\parindent=0pt
\parskip=12pt
\centerline{The histories interpretation:}
\vskip12pt
\centerline{stability instead of consistency?}
\vskip 2cm
\centerline{C J S Clarke,}

\centerline{Faculty of Mathematical Studies,}
\centerline{University of Southampton,}
\centerline{Southampton, SO17 1BJ, UK }
\vskip 2cm
{\it Abstract}
Using a reformulation of conventional results in decoherence theory, a condition is proposed for singling out a distinguished class of histories which includes those which use the ``pointer basis'' of Zurek.
\vskip 2cm
{\bf 1. What's good and bad about consistent histories}
\smallskip
The basic problem faced by any extension of quantum theory outside the laboratory realm of Bohr's interpretation can be expressed by the slogan (Giulini et al., 1996) ``how does the classical world emerge?'' Unpacking the meaning of ``classical world'' we can extract several interlinked problems:
\par
{\parindent=40pt
\item{1.} How is it that the logic of propositions about macroscopic objects is classical (that is, a Boolean lattice) whereas the logic of quantum propositions is an orthocomplemented lattice (Beltrametti \& Cassinelli, 1981)?
\item{2.} Where the classical behaviour is, as a consequence of underlying quantum processes, indeterministic, then
\itemitem{(a)} what determines the particular Boolean sub-lattice of quantum states that can be expressed as corresponding classical states (why does the Schr\"odinger's cat Gedankenexperiment have the result that we believe it would have)?
\itemitem{(b)} how is the statistical mechanics of these probabilistically chosen classical states derived from the dynamics of quantum theory?
\itemitem{(c)} on each individual occurrence, what determines which one of the classically allowed states is in fact actualised?
\item{3.} When the classical behaviour is (at least to a high accuracy) deterministic, how is the classical dynamics derived from the quantum dynamics?
\par}

I would claim that, apart from Bohm's interpretation, which relies on intrinsically unobservable hidden variables, and interpretations involving essentially new and untested physics, existing interpretations can solve 2b and 3 in many situations, but can only partially solve the others. The consistent histories approach (Omn\`es, 1999a; Giulini et al., 1996) has proceeded futhest with resolving these problems, and has done so in a way that introduces the minimum of controversional ontological scafolding (many worlds etc). I want to propose a modification of the histories approach that solves 1 and 2a more satisfactorily. 2c remains a problem for me, and for versions of the histories approach other than that of (Omn\`es, 1999b), who deconstructs the question on metaphysical grounds.

Attempts to solve 1 and 2a rest (as does 2b) on the phenomenon of decoherence which ensures that a statistical ensemble of macroscopic systems linked to microscopic states, all initially prepared in the same state, will as a result of environmental influences, evolve to an ensemble described by a density matrix that is almost exactly diagonal in a basis (the  pointer basis, Zurek, 1981; Giulini et al., 1996) adapted to the distinct classical states of the macroscopic system. This leaves two unresolved difficulties, however, which I will justify shortly:
\par
{\parindent=40pt
\item{A.} The consistency condition used by the histories programme (see (2) below) in fact implies, tautologously without any physics, that the histories obey classical logic (in the sense of satisfying probability sum rules: see Dowker and Kent (1996) eqn (2.5)). Thus 1 is not being solved by a physical explanation, but is in effect just being put in ``by hand.'' 
\item{B.} Even if we accept the consistency condition, which forces a Boolean lattice, how do we know that there is not some other Boolean lattice, in addition to that defined by macroscopically distinguishable states, which still satisfies the consistency condition? If that were the case, then the consistency condition would not determine the lattice of classical states and 2a would not be solved.
\par
}
Both these problems are of physical interest; for, if we could clearly articulate physical conditions under which a unique Boolean lattice emerged in the quantum limit, then it would become of great interest, and would be theoretically grounded, to look for areas where there were slight departures from a Boolean lattice. Some aspects of biology might provide evidence of such areas (Ho, 1998).
Point B rests (i) on the circumstances that no rigorous mathematical proof exists of the uniqueness of the pointer basis in generating a classical logic, and (ii) on the demonstration by Dowker and Kent (1996) that, both in specific examples and in general on dimension-counting grounds, this basis is not unique. The latter argument is not cast-iron, since it could be that special symmetries invalidate the general dimension-counting arguments; but the onus is now on those who claim uniqueness of the pointer basis to demonstrate it rigorously.
Point A, however, provides the main focus of this paper

\medskip
{\bf 2. Consistent histories and the stability condition }
\medskip
{\it The consistent histories formalism}\smallskip

I assume here that we are dealing with conventional quantum mechanics over a given Hilbert space ${\cal H}$. The version of the histories formalism that I am using here is taken from Dowker and Kent (1996) except that my notation interchanges their sub- and superscripts. This version is not explicitly relativistic, but I do not regard this as essential for the point being made here.

A {\it history set} is a pair $\hbox{\bf H}=(\rho, (\sigma_1,\sigma_2,\ldots,\sigma_n))$ for some $n$ where 
{\parindent=40pt
\item{}$\rho$ is a density matrix (unit trace non-negative Hermitean operator on ${\cal H}$)
\item{}for each $i$, $\sigma_i=(P^{(i)}_1, \ldots, P^{(i)}_{k_i})$ with the $ P^{(i)}_j$ being projections (interpreted as Heisenberg picture operators) satisfying
\itemitem{} $ P^{(i)}_j P^{(i)}_k = \delta_{jk} P^{(i)}_j$
\itemitem{} $\sum_{j=1}^{k_i} P^{(i)}_j = 1$
\par}
Note that through these conditions we {\it are} putting in by hand the classicality of the propositions for any one instant, but we are {\it not} demanding it overall in the way that the different $\sigma_i$ relate to each other.

A {\it history} belonging to \hbox{\bf H} is a sequence $H = (P_1,\ldots,P_n)$ with $P_i \in \sigma_i$ for all $i$ and the probability of $H$ in the initial state $\rho$ is given by
$$
\hbox{\bf P}(\rho;H) := \hbox{Tr}(P_n\ldots P_1\rho P_1\ldots P_n).\eqno(1)
$$

The {\it consistency condition} on {\bf H} in its strongest form (the arguments given above also apply to many of the weaker forms) is that 
$$
\hbox{Tr}(P^{(n)}_{i_n}\ldots P^{(1)}_{i_1}\rho P^{(1)}_{j_1}\ldots P^{(n)}_{j_n}) = \delta_{i_1j_1}\ldots\delta_{i_nj_n}\hbox{\bf P}(\rho;( P^{(1)}_{i_1},\ldots P^{(n)}_{j_n})) .\eqno(2)
$$
\medskip
{\it The stability condition}
\smallskip
This condition is based on the well known distinction (see, for example, the survey in Tegmark, 2000) between the dynamical timescale $t_d$ and the decoherence timescale $t_{dc}$ for a system. The dynamical time scale is determined by the {\it system} Hamiltonian, independently of the environment; whereas both are involved in decoherence. Dynamical timescales can vary widely: for human experience, based on neuronal firing rates, this might be $10^{-3}$s, for cosmology in the present era $10^{15}$s and for electon-positron pair production (where the logic would be highly non-classical) $10^{-20}$s. The principle of the stability condition is that the probabilities for histories should not vary (as a function of the timing of their propositions) on a timescale (the stability timescale) that is very much less than the dynamical timescale $t_d$. We cannot, for instance, claim to be talking about human experience and then introduce a proposition whose probability changes on a timescale of $10^{-10}$s. It is straight forward (see \S 3 below) to see that the probabilities for ``unphysical'' propositions (referring, for example, to superposed states of the human brain) vary on the decoherence timescale, and so this condition does precisely what is required. 

This proposal still has an air of the ad hoc about it, and needs to be related to a more fundamental theory. But it is, unlike the consistency condition, non-trivial (in the sense of point A, that it does not beg the question it is trying to solve) and is sufficiently grounded physically to point the way to a correct fundamental theory. The rest of the paper is devoted to spelling out in more detail how this operates in practice.

We consider a situation where we are examining the effect of a proposition $P$ posed after a subhistory $H^{(i)}=(P_1,\ldots,P_i)$ with an initial state of $\rho$. Thus we are concerned with $\hbox{\bf P}(\rho;H^{(i)},P)$. Now let $P_t$ denote the proposition obtained by evolving $P$ for time $t$, namely
$$
P_t := \exp (iHt/\hbar)P \exp (-iHt/\hbar).\eqno(3)
$$
Then we define the {\it repetition probability}  by
$$
p(t):= \hbox{\bf P}(\rho;H^{(i)},P,P_t).\eqno(4)
$$
Note that $p(0) > p(t)$ for $t>0$. 

We can now define the (repetition) {\it stability timescale} $t_s$ for $P$ in this context. The idea is elementary but its formulation rather tedious; again, an indication that the theory cannot be in any sense fundamental.
We want to define the stability timescale as the inverse of the gradient of $p(t)$ near $t=0$ (more precisely: the slope of the chord from $t=0$ to a suitable point). Unfortunately, $p(t)$ may be subject to small fluctuations due to perturbations from background noise (of a normal physical kind unconnected with decoherence) which could give rise to large gradients on a very small timescale. We define the magnitude of these possible fluctuations away from $t=0$ by setting
$$
V(t,c) := \sup_{{t\leq t_1 < t_2 \leq t_d} \atop {t_2 - t_1 \geq c}}
{|p(t_2) - p(t_1)|\over t_2 - t_1}   \eqno(5)
$$
Then let $F$ be the ratio of the slope of the chord from $t=0$ to the slope of the following fluctuations:
$$
F(t) := {p(0) - p(t)\over t }\Big/ V(t,t)  \eqno(6)
$$
having a supremum of $F^*$, and let $t^*$ be the smallest point (it exists!) for which 
$$\limsup_{t\to t^*}F(t) = F^*.   \eqno (7)$$ 
The stability timescale is then the inverse slope of the chord to $t^*$:
$$
t_s := {t^*\over p(0) - p(t^*)}. \eqno(8)
$$
The stability condition is then the requirement on each $\sigma_i$ that $t_s > \lambda t_d$ where $\lambda$ is some chosen small parameter.

\medskip
{\bf 3. Stability and decoherence }
\smallskip

This section fills in the obvious connection between stability and decoherence, showing that as a result of the latter, the stabiity condition rules out superpositions of macroscopically distinct states, and thus gives rise to a Boolean lattice. Decoherence involves the setting where ${\cal H} = {\cal H}_E \otimes {\cal H}_S$ where $S$ refers to the system and $E$ to the environment. It is hard to give a completely general formulation of the results concerning decoherence and the pointer basis (see Giulini et al., 1996); but a model of the idea sufficient for our purposes might be the proposition that each $\sigma_i $ can be chosen so that
{\parindent=40pt 
\item{(a)} For all $k$, $\hbox{Range} P^{(i)}_k  = {\cal H}_E \otimes V^{(i)}_k $ for a subspace $ V^{(i)}_k $ of ${\cal H}_S $
\item{(b)} If $|e\rangle \in V^{(i)}_k $ and $|f\rangle \in V^{(i)}_l $ for $k\neq l$, then $\langle e|\rho_S|f\rangle \to 0$ in the decoherence timescale, where $\rho_S$ is the density matrix projected to ${\cal H}_S$ by tracing over environment variables.
\par}
Consider, then, the possibility of measuring a projection on a superposition  $|k\rangle = a|e\rangle + b|f\rangle$ (i.e.\ a  ``live cat $+$ dead cat'' situation) with $|a|^2 + |b|^2 = 1$. Thus let $P = |k\rangle \langle k|$. Let $\rho' = H^{(i)T}\rho H^{(i)}$ where ${}^T$ denotes transpose. Then from (4) 
$$
p(t) = \hbox{Tr} P_tP\rho' PP_t.  \eqno (9)
$$
The effective density matrix following $P$, projected onto the system variables, is
$$
\rho^* := \left({P\rho'P\over p(0)}\right)_{\! S}. \eqno(10)
$$
This is a unit trace matrix proportional to $|k\rangle \langle k|$, and hence is equal to $|k\rangle \langle k|$ (${}=P$), the non-zero terms, in a basis containing $|e\rangle$ and $|f\rangle$ being
$$
\rho^* \sim \left[\matrix{|a|^2&a\bar b\cr  \bar a b&|b|^2} \right]. \eqno (11)
$$
The off-diagonal terms decay with the decoherence time $t_{dc}$, while the diagonal terms are stable and so from (9) and (10), noting that tracing over the environment commutes with $P$ (but not with $H$)
$$
p(t) =  p(0) [\matrix{a&b\cr}]\left[\matrix{|a|^2&\epsilon\cr \epsilon &|b|^2} \right]\left[\matrix{\bar a\cr \bar b\cr}\right] \eqno (12)
$$
where $|\epsilon | \sim e^{-t/t_{dc}}$.
Thus 
$$
p(t) \to p(0)(|a|^4 + |b|^4) = p(0)(1 - 2|a|^2|b|^2).
$$
This will violate the stability condition unless either $|a|$ or $|b|$ is very small, that is, unless the superposition is very close to a pure macroscopic state, as required.

\medskip
{\bf 4. Conclusion }
\smallskip
I have stressed that this is a provisional theory with ad hoc elements. There seems to be a growing feeling among some workers (e.g.\ Zeh, 2000) that the inadequacies of the current situation can only be overcome by a theory of mind; a view that I would endorse, though without thereby endorsing a many-minds metaphysics. There remains, however, a considerable gap between approaches to a theory of mind starting from the requirements of quantum theory (e.g.\ Donald, 1995) and those starting from psychology (e.g.\ Velmans, 2000). Work under way to bridge this gap (e.g. Clarke, 2000) still needs to develop an adequate dynamics; but the requirement of a stability condition can now be used to provide a clear goal in this work.

\medskip
{\bf References }
\smallskip

Beltrametti, Enrico G \& Cassinelli, Gianni (1981) {\it The Logic of Quantum Mechanics}, Addison Wesley, Reading, Mass.
\smallskip
Clarke, C J S (2000) ``Consciousness and non-hierarchical physics'' in  {\it The physical nature of consciousness,}  Ed.  Philip van Looke, Jon Benjamins Publishing
\smallskip 
Donald, M (1995) ``A mathematical characterisation of the physical structure of observers, {\it Foundations of Physics,} {\bf  25}, 529-571
\smallskip
Dowker, F  and Kent, A (1996) ``On the consistent histories approach to quantum mechanics'' {\it J. Stat. Phys.} 82 1575
\smallskip 
Omn\`es, Roland (1999a) {\it Understanding Quantum Mechanics} Princeton University Press, Princeton NJ
\smallskip 
Omn\`es, Roland (1999b) {\it Quantum philosophy} Princeton University Press, Princeton NJ
\smallskip 
Giulini, D, Joos, E, Kiefer, C, Kupsch, J, Stamatescu, I-O \& Zeh, H D (Eds) (1996) {\it Decoherence and the appearance of a classical world}, Springer-Verlag, Berlin and Heidelberg
\smallskip 
Ho, M-W (1998) {\it The Rainbow and the Worm} (2nd Edition) London: World Scientific
\smallskip
Tegmark, M (2000) {\it Phys.\ Rev.\ E} {\bf 61} (4) 4194--4206
\smallskip
Velmans, M (2000) {\it Understanding Consciousness} Routledge
\smallskip 
Zeh, H. D. (2000) ``The problem of conscious observation in quantum mechanical description'', {\it Found.\ Phys.\ Lett.} {\bf 13} 221-233
\smallskip 
Zurek, W. H. (1981) {\it Phys.\ Rev.\ D}, {\bf 24} 1516
\end